\documentclass[twocolumn]{aastex62}
\usepackage{apjfonts,amsmath,verbatim}

\shorttitle{GWs from Close Binaries with Time-Varying Masses}
\shortauthors{Holgado and Ricker}

\begin{document}

\title{GRAVITATIONAL RADIATION FROM CLOSE BINARIES WITH TIME-VARYING MASSES}

\correspondingauthor{A.~Miguel Holgado}
\email{holgado2@illinois.edu}

\author[0000-0003-4143-8132]{A.~Miguel Holgado}
\author[0000-0002-5294-0630]{Paul M. Ricker}
\affil{Department of Astronomy and National Center for Supercomputing Applications, University of Illinois at Urbana-Champaign, Urbana IL, 61801, USA
}



\begin{abstract}
In the quadrupole approximation of General Relativity in the weak-field limit, a time-varying quadrupole moment generates gravitational radiation. 
Binary orbits are one of the main mechanisms for producing gravitational waves and are the main sources and backgrounds for gravitational-wave detectors across the multi-band spectrum. 
In this Paper, we introduce additional contributions to the gravitational radiation from close binaries that arise from time-varying masses along with those produced by orbital motion.
We derive phase-dependent formulae for these effects in the quadrupolar limit for binary point masses, which reduce to the formulae that Peters and Mathews (1963) derived when the mass of each component is taken to be constant. 
We show that gravitational radiation from mass variation can be orders of magnitude greater than that of orbital motion. 
\end{abstract}


\section{Introduction} \label{sec:intro}
Einstein's theory of General Relativity (GR) predicted the existence of gravitational waves (GWs) \citep{einstein_notitle_1916,einstein_notitle_1918}. 
The existence of GWs was later indirectly confirmed from observations of the orbital decay of a pulsar binary \citep{hulse_discovery_1975}. 
Recently, the Laser Interferometer Gravitational Wave Observatory (LIGO) has ushered in the era of GW astronomy with detections of GWs from compact-object mergers \citep{ligo_scientific_collaboration_and_virgo_collaboration_observation_2016,ligo_scientific_collaboration_and_virgo_collaboration_gw151226:_2016,ligo_scientific_and_virgo_collaboration_gw170104:_2017,ligo_scientific_collaboration_and_virgo_collaboration_gw170814:_2017,abbott_gw170608:_2017,ligo_scientific_collaboration_and_virgo_collaboration_gw170817:_2017}.
The Laser Interferometer Space Antenna (LISA) will then open up the millihertz frequency band \citep{danzmann_lisa:_1996,amaro-seoane_laser_2017}. 
The combination of ground-based and space-based detectors holds promise for advances in multi-band GW astronomy, standard-siren cosmology, and tests of GR \citep{sesana_prospects_2016}. 
At nanohertz frequencies, pulsar timing arrays are currently operating and will eventually detect the stochastic GW background from the cosmic population of inspiraling supermassive black-hole binaries \citep{arzoumanian_nanograv_2018}.
Even though a detection has yet to be made, the upper limits are providing astrophysically relevant constraints on the cosmic
supermassive black-hole binary population \citep{sesana_testing_2018,holgado_pulsar_2018}.
The decihertz band may be opened with future space-based detectors. 
The main astrophysical GW sources and GW backgrounds that these detectors across the multi-band GW spectrum aim to detect are from the orbits of compact binaries. 
In this Paper, we consider an additional effect: time-varying masses in a binary due to wind mass loss or accretion from mass transfer, for example, generate time-varying quadrupole moments that generate gravitational radiation in addition to those produced by orbital motion. 
\par
The analytic expressions for gravitational radiation from point-mass binaries in the quadrupole approximation of GR in the weak-field limit \citep{peters_gravitational_1963,peters_gravitational_1964} have been used widely in the astrophysics community for modeling the evolution of close stellar binaries, generating compact-binary distributions from binary population synthesis codes \citep[e.g.,][]{belczynski_comprehensive_2002,belczynski_compact_2008}, and estimating the strain from sub-parsec supermassive black-hole binary candidates.
In this Paper, we consider the more general treatment of the two-body problem in which the mass of the binary is time-dependent. 
This is motivated by the variety of binary systems that interact, i.e., low-mass X-ray binaries, cataclysmic variables, and supernovae in close binaries.
Binary interactions include Roche-lobe mass transfer \citep[e.g.,][]{paczynski_evolutionary_1971,eggleton_approximations_1983}, mass loss, and common-envelope evolution \citep[e.g.,][]{holgado_gravitational_2018}. 
A number of previous works in the astrophysics and fundamental-physics communities have considered close binaries with time-varying masses in different contexts \citep[e.g.,][]{macedo_into_2013,barausse_can_2014,tauris_disentangling_2018}. 
\cite{cheng_orbital_2004} considered the effect of the time-varying masses on the quadrupole moment in the context of neutron-star spindown, but did not consider the phase-dependent effects that we present here. 
We derive new expressions for gravitational radiation from binaries with variable masses and we quantify the differences between treatments with and without the additional contributions for an explosive mass-loss case.
The expressions that arise from this new analysis reduce to the Peters and Mathews (PM herein) formulae \citep{peters_gravitational_1963,peters_gravitational_1964} when the mass of each component is taken to be constant.
%
\section{The Perturbed Two-Body Problem}
If two constant point masses are in an orbit bound by their mutual gravitational interaction, their motion can be described as a Keplerian orbit in the absence of any internal and external perturbations. 
The six orbital elements $(a,e,i,\varpi,\omega,\nu)$ fully define the Keplerian orbit, where $a$ is the semi-major axis, $e$ is the eccentricity, $i$ is the inclination, $\varpi$ is the longitude of the ascending node, $\omega$ is the argument of periapsis, and $\nu$ is the true anomaly.
We consider a gravitationally bound point-mass binary with time-dependent component masses $m_1(t)$ and $m_2(t)$ with a reduced mass $\mu = m_1 m_2/M(t) \equiv \mu(t)$, where $M(t) =m_1+m_2$ is the total mass.
Allowing for time-dependent masses introduces perturbative forces on the two bodies. 
\par
We consider the case where the primary's mass remains constant and the secondary undergoes isotropic mass loss: $\dot{m_1} = 0$ and $\dot{m}_2 < 0$. 
The resulting total mass-loss rate of the binary is $\dot{M} = \dot{m}_2$. 
For the isotropic case, only the self-gravity between the two components governs the equation of motion for the relative separation vector ${\bf r}$
\begin{equation} \label{eq:eom}
\ddot{\bf r} = - \frac{GM}{r^3} {\bf r}  \ ,
\end{equation}
where $G$ is the gravitational constant, $r$ is the magnitude of ${\bf r}$, and where we take the same coordinate system as \cite{peters_gravitational_1963} with the origin located at the barycenter.
There is an implicit perturbing force that induces the following acceleration
\begin{equation}
{\bf f} = -\frac{1}{2} \frac{\dot M}{M}\dot{\bf r} \ ,
\end{equation}
which is implicit in the isotropic variation of the mass-losing body. 
This has been derived in several previous works \citep{hadjidemetriou_two-body_1963,hadjidemetriou_analytic_1966,hadjidemetriou_binary_1966,verhulst_asymptotic_1975}. 
If the mass variation is anisotropic, i.e., either from non-conservative mass exchange and asymmetric ejecta, then the equation of motion will have an explicit perturbing force \citep[e.g.,][]{dosopoulou_orbital_2016-1,dosopoulou_orbital_2016}. 
The orbital evolution can be described as an osculating orbit, which is the Keplerian orbit the binary would follow if the perturbation ceased immediately.
The full set of equations for the time evolution of the osculating orbital elements from the isotropically varying mass are \citep[e.g.,][]{hadjidemetriou_two-body_1963}
\begin{subequations}
\begin{align}
\frac{{\rm d}a}{{\rm d}t} &= - \left(\frac{1+2e\cos\nu + e^2}{1-e^2}\right) \frac{\dot M}{M} a \ , \\
\frac{{\rm d}e}{{\rm d}t} &= - (e+\cos\nu) \frac{\dot M}{M} \ , \\
\frac{\rm d\omega}{{\rm d}t} &= -\frac{\sin \nu}{e} \frac{\dot M}{M} \label{eq:precess} \ , \\
\frac{\rm d\nu}{{\rm d}t} &= \frac{\left[G M a \left(1-e^2\right)\right]^{1/2}(1+e\cos\nu)^2}{a^2 \left(1-e^2\right)^2} + \frac{\sin \nu}{e} \frac{\dot M}{M} \ ,
\end{align}
\end{subequations}
where the specific angular momentum is conserved, i.e., the quantity $r^2 \dot{\theta} = \left[G M a \left(1-e^2\right)\right]^{1/2}$ is a constant of the motion. 
Since there is no net force in the direction perpendicular to the orbital plane, $\dot{i} = \dot{\varpi} = 0$, and the orbital motion thus remains coplanar. 
The mass variation causes the argument of periapsis to precess at a rate $\dot{\omega}$ (see Eq.~\ref{eq:precess}).
%
\section{Quadrupole Gravitational Radiation}
For eccentric Keplerian orbits, the orbital separation $r$ between the binary masses obeys
\begin{equation}
r = \frac{a\left(1-e^2\right)}{1+e\cos\nu} \ .
\end{equation}
The components of the mass quadrupole tensor $I_{jk}$ for an eccentric binary are
\begin{subequations}
\begin{align}
I_{xx} &= \mu r^2 \cos^2 \theta = \mu a^2 \left(1-e^2\right)^2 \frac{\cos^2\theta}{(1+e\cos\nu)^2} \ , \\
I_{yy} &= \mu r^2 \sin^2\theta = \mu a^2 \left(1-e^2\right)^2 \frac{\sin^2\theta}{(1+e\cos\nu)^2} \ , \\
I_{xy} &= \mu r^2 \sin \theta \cos \theta = \mu a^2 \left(1-e^2\right)^2 \frac{\sin\theta\cos\theta}{(1+e\cos\nu)^2} \ ,
\end{align}
\end{subequations} 
where $\theta = \nu + \omega$, which simplifies to $\theta = \nu$ if the precession of the argument of periapsis is taken to be negligible.
We can choose the initial $\theta$ to begin where $\omega = 0$.  
For isotropic wind mass loss, the periapsis distance $r_{\rm p} = a(1-e)$ is constant, so ${\dot r}_{\rm p} = 0$.
When including gravitational radiation reaction, the periapsis distance will indeed decrease, though one can take ${\dot r}_{\rm p} \approx 0$ on timescales less than the mass-loss timescale $M/|\dot{M}|$ when the GW coalescence timescale is much longer than the mass-loss timescale. 
The third time derivatives of the mass-quadrupole tensor components are
\begin{widetext}
\begin{subequations} \label{eq:ecc}
\begin{align}
\dddot{I}_{xx} &= - \frac{2 G \mu M}{r^2} \dot{r} \cos^2\theta + \frac{4 G \mu M}{r} \dot{\theta} \sin 2\theta - \frac{2 G \mu \dot{M}}{r} \cos^2\theta - \frac{6 G \dot{\mu} M}{r} \cos^2\theta + 6 \dot{\mu} r^2 \dot{\theta}^2 \left(\frac{e^2 \sin^2\nu \cos^2 \theta}{(1+e\cos\nu)^2} - \frac{e \sin \nu \sin 2\theta}{1+e\cos\nu} + \sin^2\theta\right) \\ \nonumber &\phantom{=}+ 3 \ddot{\mu} r^2 \dot{\theta} \left(\frac{2 e \sin \nu \cos^2\theta}{1 + e\cos\nu} - \sin 2\theta\right) + \dddot{\mu} r^2 \cos^2\theta \ , \\
\dddot{I}_{yy} &= - \frac{2G\mu M}{r^2} \dot{r} \sin^2\theta - \frac{4 G \mu M}{r} \dot{\theta} \sin 2\theta - \frac{2 G \mu \dot{M}}{r} \sin^2\theta - \frac{6 G \dot{\mu} M}{r} \sin^2 \theta + 2 \dot{\mu} r^2 \dot{\theta}^2 \left(\frac{e^2 \sin^2 \nu \sin^2\theta}{(1+e\cos\nu)^2} + \frac{e\sin\nu\sin 2\theta}{1+e\cos\nu} + \cos^2\theta\right) \\ \nonumber &\phantom{=}+ 3 \ddot{\mu} r^2 \dot{\theta} \left(\frac{2 e\sin\nu \sin^2\theta}{1+e\cos\nu} - \sin 2\theta\right) + \dddot{\mu} r^2 \sin^2 \theta \ , \\
\dddot{I}_{xy} &= - \frac{G\mu M}{r^2} \dot{r} \sin 2\theta - \frac{4 G \mu M}{r} \dot{\theta} \cos 2\theta - \frac{G \mu \dot{M}}{r} \sin 2\theta - \frac{3 G \dot{\mu} M}{r} \sin 2\theta + 3 \mu r^2 \dot{\theta} \left(\frac{e^2 \sin^2 \nu \sin 2\theta}{(1+e\cos\nu)^2} + \frac{2 e \sin \nu \cos 2\theta}{1+e\cos\nu} - \sin 2\theta\right) \\ \nonumber &\phantom{=}+ 3 \ddot{\mu} r^2 \dot{\theta} \left(\frac{e \sin \nu \sin 2\theta}{1+e\cos\nu} + \cos 2\theta\right) + \frac{1}{2}\dddot{\mu} r^2 \sin 2\theta \ .
\end{align}
\end{subequations}
\end{widetext}
If the periapsis precession is taken to be negligible, $\dot{\omega} = 0$, then the first two terms of Eqs.~\eqref{eq:ecc} simplify to the canonical PM expressions. 
The additional terms in Eqs.~\eqref{eq:ecc} contain time derivatives of the masses of the binary, which may also be thought of as time derivatives of the binary's chirp mass ${\cal M} = (m_1 m_2)^{3/5}/ M^{1/5}$. 
\par
Gravitational radiation carries away the orbital energy and angular momentum of the binary at the following rates
\begin{subequations}
\begin{align} \label{eq:power}
\left. \frac{{\rm d}E}{{\rm d}t}\right|_{\rm GW} &= \frac{1}{5} \frac{G}{c^5} \left\langle \dddot{\cal I}_{jk} \dddot{\cal I}_{jk} \right\rangle \ , \\
\left. \frac{{\rm d}J_i}{{\rm d}t}\right|_{\rm GW} &= \frac{2}{5} \frac{G}{c^5} \epsilon_{ijk}\left\langle \ddot{\cal I}_{jm} \dddot{\cal I}_{km} \right\rangle \ ,
\end{align}
\end{subequations}
where $c$ is the speed of light and ${\cal I}_{jk}$ is the reduced (trace-free) quadrupole moment tensor ${\cal I}_{jk} = I_{jk} - \frac{1}{3} \delta_{jk} \delta_{\ell m} I_{\ell m}$.
%
%
If we take the orbit to be circular, then the GW luminosity simplifies to
\begin{multline} \label{eq:lum}
L_{\rm GW} = \frac{32}{5} \frac{G^4}{c^5} \frac{\mu^2 M^3}{a^5} + \frac{72}{5} \frac{G^3}{c^5} \frac{{\dot \mu}^2 M^2}{a^2} - \frac{48}{5} \frac{G^3}{c^5} \frac{\ddot{\mu} \mu M^2 }{a^2} \\ 
- \frac{12}{5} \frac{G^2}{c^5} \dddot{\mu} \dot{\mu} M a + \frac{18}{5} \frac{G^2}{c^5} \ddot{\mu}^2 M a + \frac{1}{5}\frac{G}{c^5} \dddot{\mu}^2 a^4 \ ,  
\end{multline}
where the positive terms in Eq.~\eqref{eq:lum} correspond to GW energy loss from the orbit and the negative terms correspond to GW energy gain in the orbit.
From the strain tensor in the transverse-traceless gauge,
\begin{equation} \label{eq:strain}
h_{jk}^{\rm TT} = \frac{2 G}{c^4 D_{\rm L}} \ddot{\cal I}_{jk} \ ,
\end{equation}
where $D_{\rm L}$ is the luminosity distance to the source, the strain amplitude for a circular binary is
\begin{multline} \label{eq:strain}
h_0 = \frac{4 G^{5/3}}{c^4} \frac{\mu M^{2/3}\Omega^{2/3}}{D_{\rm L}} + \frac{2 G^{5/3}}{c^4} \frac{{\dot \mu} M^{2/3}}{D_{\rm L} \Omega^{1/3}} + \frac{G^{5/3}}{c^4} \frac{\ddot{\mu} M^{2/3}}{D_{\rm L}\Omega^{4/3}} \ ,
\end{multline}
where $\Omega$ is the angular frequency of the binary, $\Omega = 2 \pi f_{\rm orb} = 2\pi / P_{\rm orb}$.
In the transverse-traceless gauge, the plus and cross polarizations of the strain are $h_+ = h_{xx}^{\rm TT} = - h_{yy}^{\rm TT}$ and $h_\times = h_{xy}^{\rm TT} = h_{yx}^{\rm TT}$.
From Eq.~\eqref{eq:strain}, we can define the following parameters
\begin{subequations}
\begin{align}
\zeta_1 \equiv \frac{\dot{\mu}/\mu}{2\Omega} \ , \\
\zeta_2 \equiv \frac{\ddot{\mu}/\mu}{\left(2\Omega\right)^2} \ ,
\end{align}
\end{subequations}
which quantify the relative importance of mass-variation effects on the GW amplitude relative to the PM term of a given source.
For $\zeta_1 = 1$, the mass-transfer timescale needs to be comparable to the orbital period.
The true trajectory may differ significantly from that of an osculating orbit since the perturbing force in this regime is comparable to the self-gravity. 
\par
We emphasize that even though the above analysis is solely applicable to a binary whose mass is varying isotropically, it is illustrative of how one includes time-varying masses in the PM analysis and eventually extensions to the post-Newtonian analysis. 
In the limit that the time-derivatives of the masses are zero, then the canonical PM expressions remain. 
\section{Explosive Mass Loss in a Close Binary}
Supernova explosions in close binaries result in extreme mass loss. 
If enough mass is ejected during the explosion, then the binary can be disrupted, i.e., the final energy of the binary is $\ge 0$ and the orbit becomes unbound. 
We again consider the case of isotropic mass loss such that the kick imparted to the remnant of the exploding body is solely due to the acceleration from the ejected mass (i.e., a ``Blaauw kick'' \citep{blaauw_origin_1961}), such that the equation of motion remains the same as Eq.~\eqref{eq:eom}. 
In this regime, we choose to work in Cartesian coordinates such that ${\bf r} = (x,y,z)$, $\dot{\bf r} = (\dot{x},\dot{y},\dot{z})$, and $\ddot{\bf r} = (\ddot{x}, \ddot{y}, \ddot{z})$.
We also ignore any interaction between the companion and the ejecta. 
When the mass-loss timescale is shorter than the orbital period, the perturbing force can become greater than the gravitational attractive force and the motion can no longer be accurately described as an osculating Keplerian orbit.
One then needs to integrate the equation of motion directly. 
The mass quadrupole tensor of the binary in these coordinates is  
\begin{equation}
I_{jk} =
\begin{pmatrix}
\mu x^2 & \mu xy & \mu xz \\
\mu xy & \mu y^2 & \mu yz \\
\mu xz & \mu yz & \mu z^2
\end{pmatrix}
\ ,
\end{equation}
and the third derivatives of each component are
%
\begin{subequations} \label{eq:general}
\begin{align}
\dddot{I}_{xx} &= \dddot{\mu} x^2 + 6 \ddot{\mu} x \dot{x} + 6 \dot{\mu} \dot{x}^2 + 6 \dot{\mu} x \ddot{x} + 6 \mu \dot{x} \ddot{x} + 2 \mu x \dddot{x} \ , \\
\dddot{I}_{yy} &= \dddot{\mu} y^2 + 6 \ddot{\mu} y \dot{y} + 6 \dot{\mu} \dot{y}^2 + 6 \dot{\mu} y \ddot{y} + 6 \mu \dot{y} \ddot{y} + 2 \mu y \dddot{y} \ , \\
\dddot{I}_{zz} &= \dddot{\mu} z^2 + 6 \ddot{\mu} z \dot{z} + 6 \dot{\mu} \dot{z}^2 + 6 \dot{\mu} z \ddot{z} + 6 \mu \dot{z} \ddot{z} + 2 \mu z \dddot{z} \ , \\
\dddot{I}_{xy} &= \dddot{I}_{yx} = \dddot{\mu} xy + 3 \ddot{\mu} \dot{x} y + 3 \dot{\mu} \ddot{x} y + \mu \dddot{x} y + 3 \ddot{\mu} x \dot{y} \\ \nonumber &\phantom{yx = =} + 6 \dot{\mu} \dot{x} \dot{y} + 3 \mu \ddot{x} \dot{y} + 3 \dot{\mu} x \ddot{y} + 3 \mu \dot{x} \ddot{y} + \mu x \dddot{y} \ , \\
\dddot{I}_{xz} &= \dddot{I}_{zx} = \dddot{\mu} xz + 3 \ddot{\mu} \dot{x} z + 3 \dot{\mu} \ddot{x} z + \mu \dddot{x} z + 3 \ddot{\mu} x \dot{z} \\ \nonumber &\phantom{zx = =} + 6 \dot{\mu} \dot{x} \dot{z} + 3 \mu \ddot{x} \dot{z} + 3 \dot{\mu} x \ddot{z} + 3 \mu \dot{x} \ddot{z} + \mu x \dddot{z} \ , \\
\dddot{I}_{yz} &= \dddot{I}_{zy} = \dddot{\mu} yz + 3 \ddot{\mu} \dot{y} z + 3 \dot{\mu} \ddot{y} z + \mu \dddot{y} z + 3 \ddot{\mu} y \dot{z} \\ \nonumber &\phantom{zy = =} + 6 \dot{\mu} \dot{y} \dot{z} + 3 \mu \ddot{y} \dot{z} + 3 \dot{\mu} y \ddot{z} + 3 \mu \dot{y} \ddot{z} + \mu y \dddot{z} \ .
\end{align}
\end{subequations}
%
The GW luminosity and strain waveform can then be calculated with these expressions using Eqs.~\eqref{eq:power} and \eqref{eq:strain}, respectively. 
Eqs.~\eqref{eq:general} are general for any point-mass binary with a given equation of motion and time-dependent masses.
These equations can also be translated into any choice of coordinate system. 
Since there are no forces acting in the $z$ direction, then the $z$ components will be zero and the orbital motion will remain coplanar in the $x$-$y$ plane. 
\par
In binary population synthesis codes, supernova mass loss in close binaries is often treated as being effectively instantaneous and conservative of the total orbital energy and orbital angular momentum. 
When treating the gravitational radiation, the mass-loss timescale should be bounded by the light-crossing timescale of the binary.
The total energy and total angular momentum can be treated as conserved as long as the amount of energy and angular momentum radiated away in GWs is included in the balance equations.
Supernovae in close binaries are the most extreme cases of mass loss and are thus expected to be the loudest sources of GWs from mass variation.
\subsection{Numerical Comparison}
We consider a close binary consisting of a neutron star and a helium star in an initially circular orbit with an initial semi-major axis of $a_0 = 10^{-2} \ {\rm au}$. 
We fix the neutron-star mass to be $m_{\rm NS,1} = 1.4 M_\odot$ and the helium-star mass to be $m_{\rm He} = 2.4 M_\odot$. 
The helium star may be large enough to be tidally deformed, but we ignore this effect and are solely interested in the GWs sourced by the supernova mass loss since this process operates on dynamical timescales, while the tidal effects on the orbit operate on secular timescales.
The helium star will undergo a supernova explosion, which we take to be isotropic for simplicity and will form a neutron star with the same mass as the primary $m_{\rm NS,2} = m_{\rm NS,1} = 1.4 M_\odot$. 
The supernova explosion itself will be a source of GWs in the LIGO band, but here, we solely focus on the GWs sourced by the binary mass loss, whose characteristic frequency corresponds to the mass-loss timescale.
We choose the following smooth function as a model for the mass-losing body such that the time derivatives of the mass are continuous:
\begin{equation} \label{eq:mass}
m_2(t) = m_{2,\rm f} + \frac{1}{2}\left(m_{2,\rm i} - m_{2, \rm f}\right) \left(1 + \frac{2}{\pi}\tan^{-1} \left[(t-t_0)/\tau\right]\right)\ ,
\end{equation}
where $m_{2,\rm i}$ is the initial mass, $m_{2,\rm f}$ is the final mass, $\tau$ is a given timescale for the mass loss and $t_0$ is the the time at which the mass loss peaks. 
Given that the only constraint for $\tau$ is the light-crossing time $\tau_{\rm c}$ of the binary, we consider two cases: $\tau = [\tau_{\rm c}, 100 \tau_{\rm c}]$. 
The light-crossing time relative to the orbital period for this particular system is $\Omega \tau_{\rm c} \approx 0.002$, so the mass-loss timescale for both cases are less than the orbital period. 
Other smooth functions may be used to model the sharp decrease in mass, but such choices will not affect our overall results. 
\par
\begin{figure}
\centering
\includegraphics[width=\columnwidth]{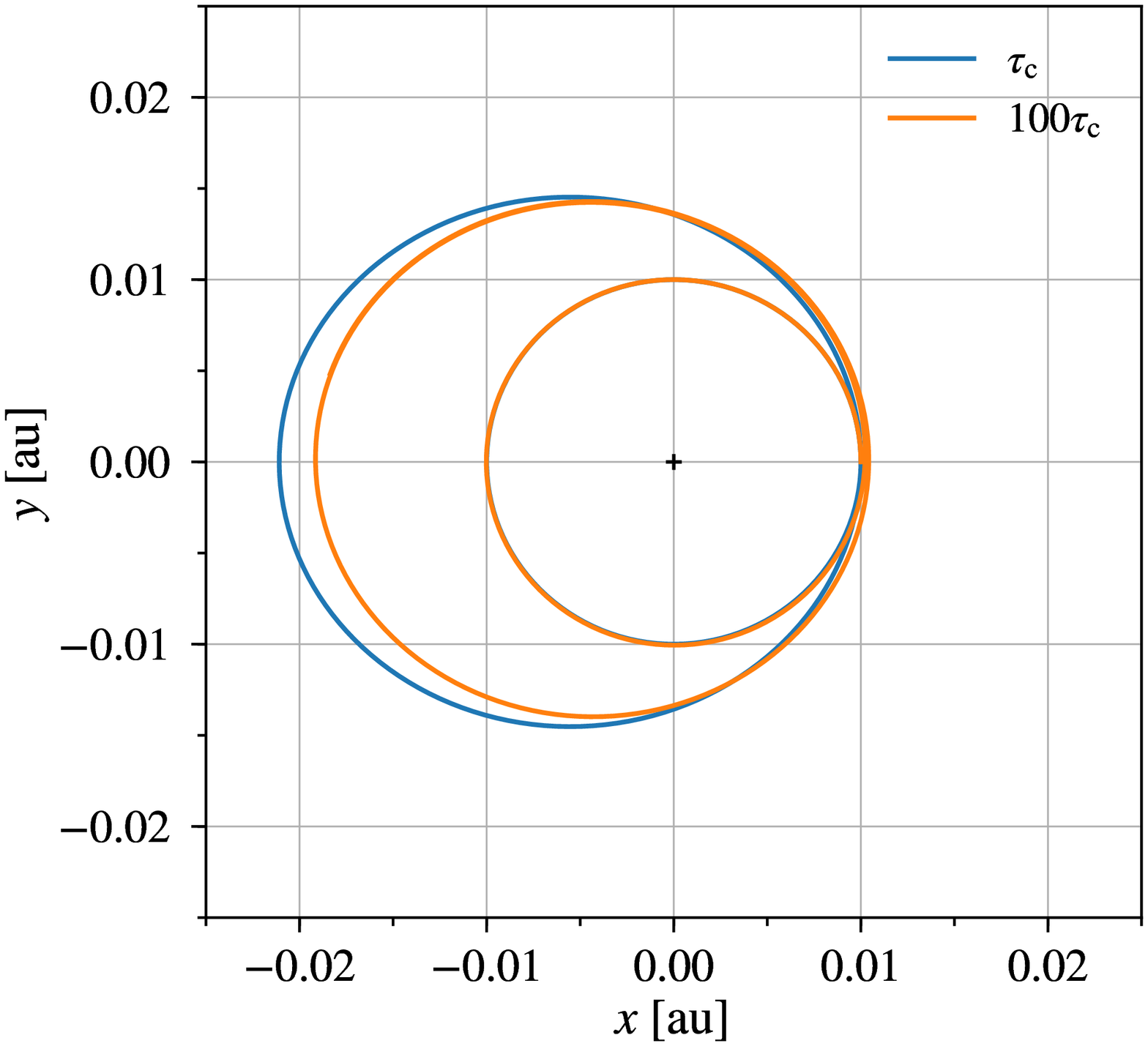}
\includegraphics[width=\columnwidth]{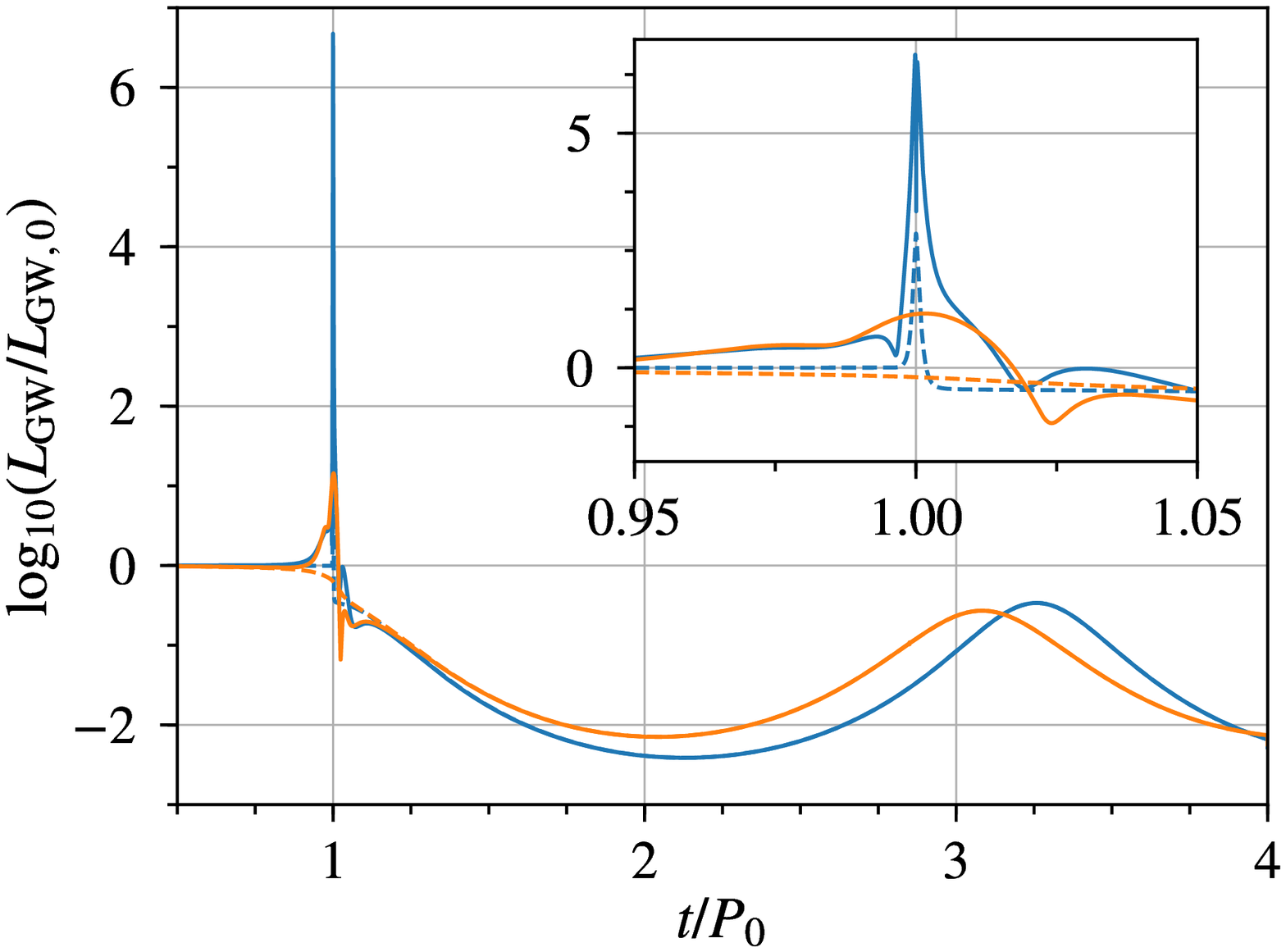}
\caption{\label{fig:compare} Top panel: orbital trajectories for the mass-loss timescales of $\tau = \tau_{\rm c}$ (blue line) and $\tau = 100 \tau_{\rm c}$ (orange line). 
The orbit is initially circular and transitions to an elliptic one as mass is lost from the binary. 
Bottom panel: GW luminosity relative to that of the initial circular orbit as the orbit evolves from its initial circular state to the final eccentric state. 
The solid lines correspond to the prediction from Eqs.~\eqref{eq:general} and the dashed lines correspond to the prediction from the PM formalism (i.e., ignoring the time-derivatives of the reduced mass in Eqs.~\eqref{eq:general}). 
An inset plot is zoomed into the region where the mass loss peaks.
The tick spacing on the $y$-axis of the inset plot is 1.25 in units of $\log_{10} \left(L_{\rm GW}/L_{\rm GW,0}\right)$.
}
\end{figure}

We plot the orbital trajectory for the two cases in the top panel Figure \ref{fig:compare}. 
For each case, we integrate the equation of motion (see Eq.~\ref{eq:eom}) with the mass of the secondary obeying Eq.~\eqref{eq:mass}.
The peak of the mass loss is taken to occur at $t_0 = P_0$, where $P_0$ is the initial orbital period of the binary (obtained from Kepler's Third Law). 
The GW luminosity evaluated from Eqs.~\eqref{eq:power} and \eqref{eq:general} and the GW luminosity predicted from the PM formalism [i.e., excluding terms in Eqs.~\eqref{eq:general} with ${\cal O} (\ge\dot{\mu})$] are plotted in the bottom panel of Figure \ref{fig:compare}. 
A spike in the GW luminosity occurs during the explosive mass loss and the amplitude of the spike increases with a decreasing mass-loss timescale.
The dashed lines in the bottom panel of Figure \ref{fig:compare} correspond to the GW luminosity computed with the PM formalism, i.e., ignoring the time-derivatives of the reduced mass in Eqs.~\eqref{eq:general}. 
The maximum difference between our predicted GW luminosity and that of PM is about 4.18 and 1.37 for the $\tau_{\rm c}$ and $100 \tau_{\rm c}$, respectively, in units of $\log_{10} \left(L_{\rm GW}/L_{\rm GW,0}\right)$.
Even for the longer mass-loss timescale case, our predicted GW luminosity is about an order of magnitude larger than the PM prediction.  
The blue dashed line, corresponding to the PM case with $\tau = \tau_{\rm c}$ exhibits a peak which comes from the acceleration on the binary from the mass loss, i.e., the Blauuw kick.
Since the mass loss is isotropic, the final orbital configuration is in the same orbital plane as the initial one. 
A more systematic and comprehensive survey of the parameter space is left for future study, especially since additional physics such as natal kicks from asymmetric supernovae will affect the GW luminosity and final state of the binary \citep[e.g.,][]{hills_effects_1983,kalogera_orbital_1996}.
\section{Discussion and Conclusions} \label{sec:conclude}
In this Paper, we have shown that time-varying masses in close binaries contribute additional gravitational radiation along with the radiation produced by orbital motion. 
We have also shown that gravitational radiation from time-varying masses can be orders of magnitude greater than that of the orbital motion.
This analysis is performed in the quadrupolar approximation of GR in the weak-field limit and the derived formulae reduce to the PM formulae when the mass of each binary component is taken to be constant. 
\par
Our analysis may be useful for implementations in binary population synthesis and generating approximate waveforms.
It may also provide new testable predictions for the evolution of compact binaries. 
Applying the analysis of this Paper to higher-order post-Newtonian orders for higher-compactness binary systems and to the gravitational self-force for high to extreme mass-ratio binaries is the next extension of this work.
These extensions may reveal richer physical behavior than the quadrupole approximation that is presented here. 
Future work will also include a more comprehensive survey of the parameter space including natal kicks from asymmetric supernovae. 
\par
Additional future studies will investigate how these new contributions to the gravitational radiation will affect the formation of compact binaries and the rates and distributions of compact-object mergers. 
The analysis we present can also be extended to hierarchical triple systems.  
A detection of GWs from mass transfer would not only validate this effect, but may also provide new and complementary tests of GR at different regimes of curvature and field strengths. 
%
\acknowledgments
AMH is supported by the DOE NNSA Stewardship Science Graduate Fellowship under grant number DE-NA0003864.
AMH thanks Charalampos Markakis, Eliu Huerta, Gabrielle Allen, Xin Liu, Charles Gammie, Nico Yunes, Fred Rasio, and Vicky Kalogera for helpful comments and fruitful discussions.
PMR acknowledges support from NSF under AST 14-13367.

%



 \software{Matplotlib \citep{hunter_matplotlib:_2007}, Numpy \citep{walt_numpy_2011} }.

\bibliographystyle{yahapj}
\bibliography{references}
\end{document}